\documentclass[letterpaper]{jpconf}
\usepackage{graphicx}
\begin{document}
\title{Toward Five-dimensional Core-collapse Supernova Simulations}

\author{C~Y~Cardall$^{1,2}$, A~O~Razoumov$^{1,2,3}$, E~Endeve$^{1,2,3}$, E~J~Lentz$^{1,2,3}$, and A~Mezzacappa$^1$}

\address{$^1$ Physics Division, Oak Ridge National Laboratory, 
Oak Ridge, TN 37831-6354, USA}
\address{$^2$ Department of Physics and Astronomy, 
University of Tennessee, 
	Knoxville, TN 37996-1200, USA}%
\address{$^3$ Joint Institute for Heavy Ion Research, 
        Oak Ridge National Laboratory, 
        Oak Ridge, TN 37831-6374, USA}

\ead{cardallcy@ornl.gov}


\begin{abstract}
The computational difficulty of six-dimensional neutrino radiation hydrodynamics has spawned a variety of approximations, provoking a long history of uncertainty in the core-collapse supernova explosion mechanism. Under the auspices of the Terascale Supernova Initiative, we are honoring the physical complexity of supernovae by meeting the computational challenge head-on, undertaking the development of a new adaptive mesh refinement code for self-gravitating, six-dimensional neutrino radiation magnetohydrodynamics. This code---called {\em GenASiS,} for {\em Gen}eral {\em A}strophysical {\em Si}mulation {\em S}ystem---is designed for modularity and extensibility of the physics. Presently in use or under development are capabilities for Newtonian self-gravity, Newtonian and special relativistic magnetohydrodynamics (with `realistic' equation of state), and special relativistic energy- and angle-dependent neutrino transport---including full treatment of the energy and angle dependence of scattering and pair interactions.
\end{abstract}

\section{Core-collapse Supernovae}

A core-collapse supernova is a colossal explosion resulting from the death of a massive star. At its peak, a supernova can outshine the combined output of the billion or so stars in its host galaxy. The star's demise is brought about by an accumulation of iron---the final product of stellar burning---in the star's core. Collapsing catastrophically, the core increases in average density by about a factor of a million in only half a second. This usually results in a hot and dense neutron star with a mass about the same as the Sun, but compressed to a size of tens of kilometers. (The result of collapse in the most extreme cases is a black hole.)

The gravitational potential energy released during collapse is initially stored as heat in the nascent neutron star, but a small fraction of this energy eventually gives rise to the spectacular disruption of the entire star. About 99\% of the energy escapes directly in the form of neutrinos, ghostly particles with very weak interactions. The explosion is caused by the remaining 1\% of this energy, which is transferred to the outer layers of the star through some yet-to-be-elucidated combination of heating by neutrinos and magnetohydrodynamic processes.

\section{Computing in Five Dimensions}

The high dimensionality of the core-collapse supernova problem is a result of the need to do neutrino transport. Inside the dense neutron star, neutrinos diffuse slowly out of the core, and their angular distribution is isotropic; but after escaping the neutron star, their angular distribution becomes more and more strongly forward-peaked. The transition between these two regimes occurs in the crucial region between the neutron star and the shock resulting from the halt of collapse of the inner core; this is the region in which neutrino heating may reenergize the stalled shock and allow the explosion to proceed. Because the heating rate depends on the energy and angle distributions of the neutrinos, one should ideally do ``neutrino radiation transport,'' in which these distributions of the neutrinos are tracked at every point in space---a six-dimensional problem.

The computational difficulty of six-dimensional neutrino radiation hydrodynamics has spawned a variety of approximations, provoking a long history of uncertainty in the core-collapse supernova explosion mechanism \cite{mezzacappa05,cardall05}. Recognizing the possible impact of convection and rotation on the efficiency and distribution of neutrino heating, some groups have focused on these while treating neutrino transport in a crude, averaged manner, seeing explosions as a result. Other groups have focused on doing more accurate neutrino transport, which required a restriction to spherical symmetry; in this case no explosions were seen. Only now are computations with three total dimensions becoming routine; soon we hope to push the envelope, with simulations in five dimensions: two space dimensions, together with a modest discretization of all three momentum space dimensions.

\section{{\em GenASiS:} A New Adaptive Mesh Refinement Code}

The computational demands of radiation hydrodynamics can be 
ameliorated by `adapative mesh refinement' (AMR). The basic idea
of AMR is to employ high resolution only where needed, in order to
conserve memory and computational effort. Our current expectation is to allow for refinement only in the space dimensions. This will help with the management of two difficulties: the large dynamic range in length scales associated with the density increase of six orders of magnitude that occurs during core collapse, and adequate resolution of particular features of the flow (the shock, for instance). With Eulerian codes in multiple space dimensions, these tasks require high resolution; and particularly for radiation hydrodynamics, the savings achievable by reducing the number of zones is considerable, since an entire three-dimensional momentum space is carried by each spatial zone. 

\begin{figure}
\includegraphics[width=4.5in,angle=270]{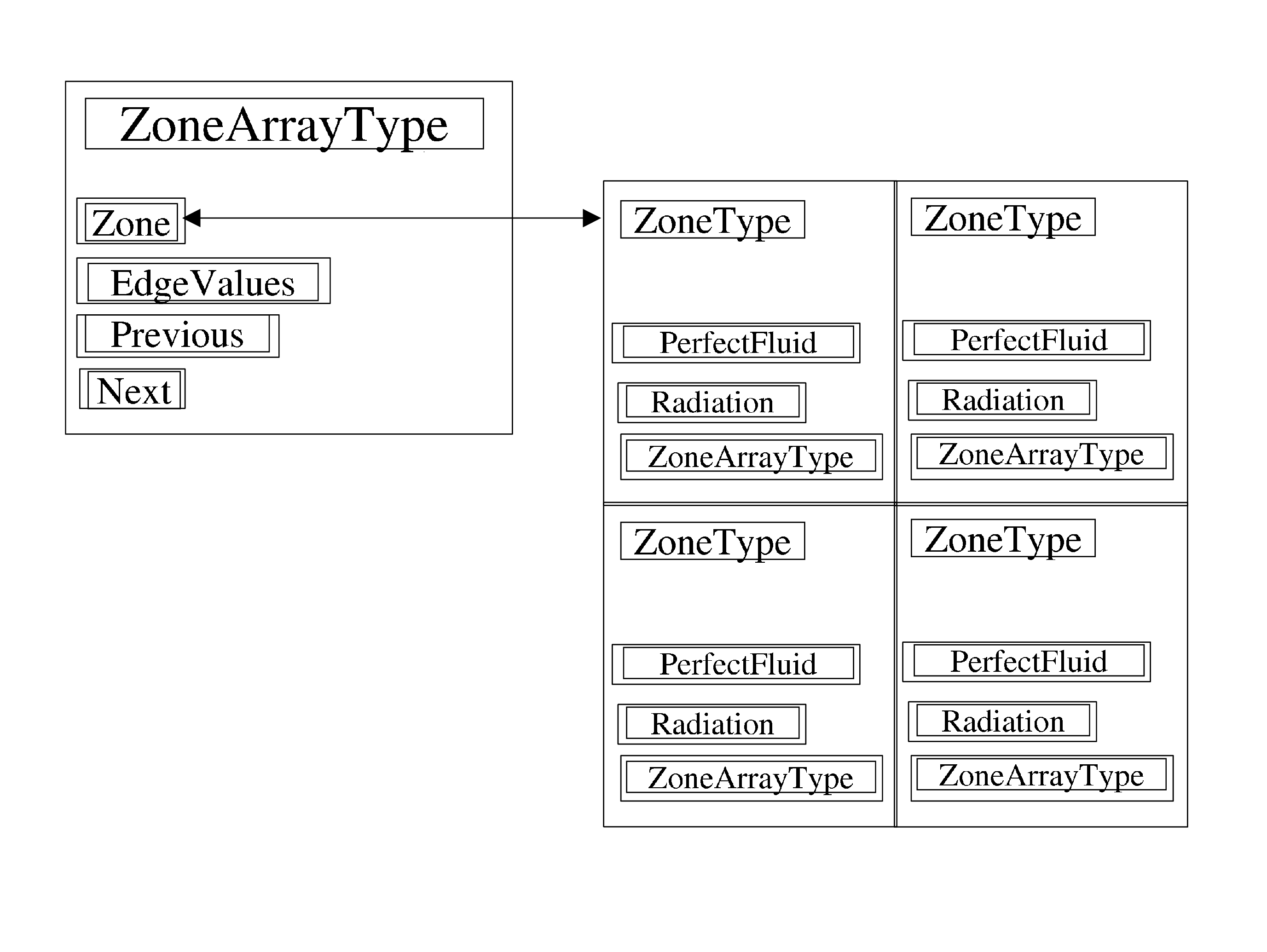}
\caption{Data structures used in an adaptive mesh for radiation 
hydrodynamics.}
\label{zone}
\end{figure}

\begin{figure}
\includegraphics[width=6.0in]{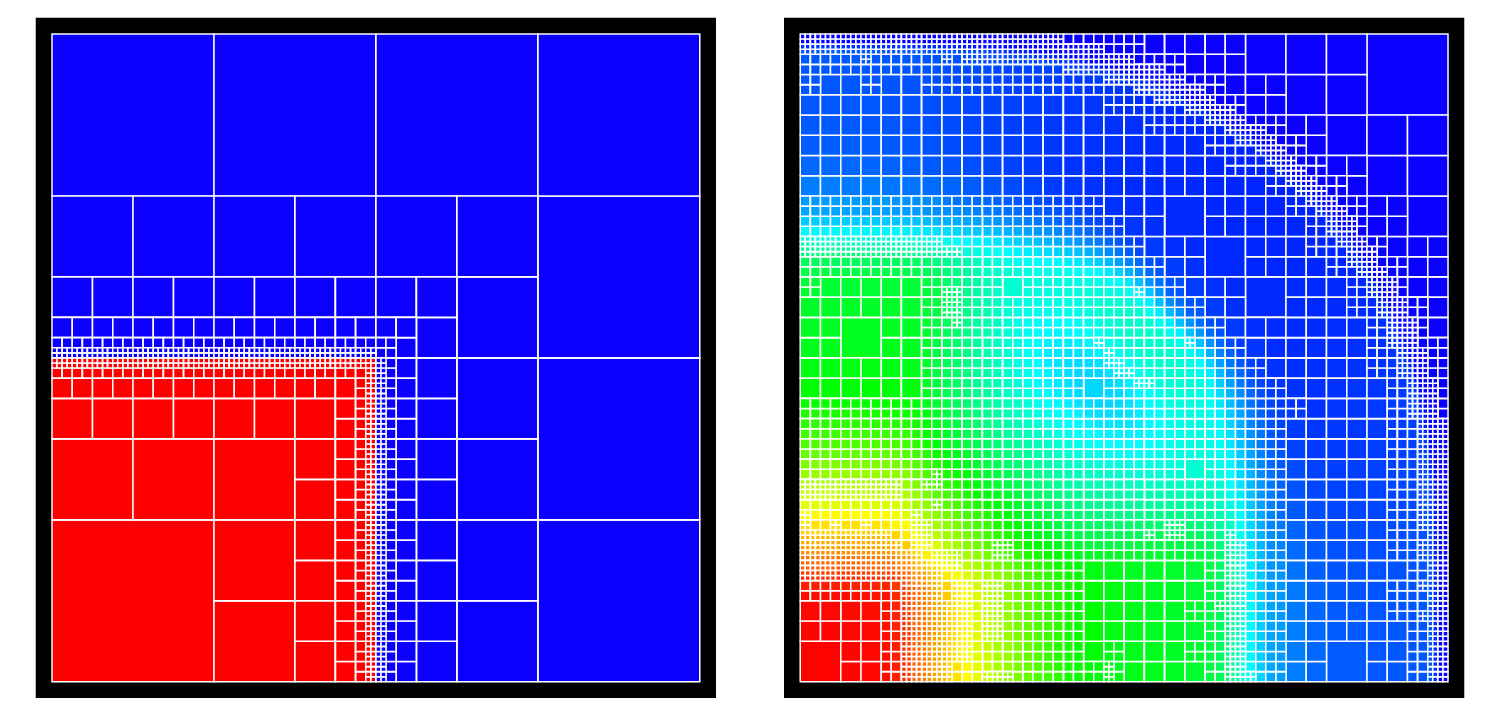}
\caption{Density in a two-dimensional generalization of the shock
tube. 
Left: Initial state, with high density in the lower left corner. Right: Evolved state.}
\label{initialFinal2D}
\end{figure}

In implementing the zone-by-zone-refinement approach to the representation of spacetime, we have tried to follow object-oriented design principles to the extent allowed by Fortran 90/95.
Figure \ref{zone}
outlines the basic data structures we use to model the ideal of a 
continuous spacelike slice with a discretized approximation. (The hierarchy of structures, and our operations on them with well-controlled interfaces, are instances of the object-oriented principles of {\em inheritance} and {\em encapsulation}.) A region
of a spacelike slice is represented by an object of {\tt zoneArrayType}. 
Each such object contains an array of objects of {\tt zoneType}, along with  
information about the coordinates of the zones and pointers to neighboring 
zone arrays. Each zone, an object of {\tt zoneType}, contains various forms
of stress-energy, each of which is a separate object. Figure \ref{zone}
shows a perfect fluid and a radiation field; in the code we have an electromagnetic field as well. 
Each zone has a pointer to another object of {\tt zoneArrayType}, whose allocation
constitutes refinement of that zone;
this structure can be extended to arbitrarily deep. A two-dimensional hydrodynamics test problem computed with our adaptive mesh code is shown in figure \ref{initialFinal2D}.

The word ``General'' that goes into the name of our code, {\em GenASiS}---for {\em Gen}eral {\em A}strophysical {\em Si}mulation {\em S}ystem---refers to the use of Fortran 90's facility for function overloading. (This is an instance of the object-oriented principle of {\em polymorphism}.) This allows a generic function name to have several different implementations, providing for extensibility of the physics: Different equations of state, hydrodynamic flux methods, coordinate systems, gravity theories, and so forth can be employed by adding new implementations of generic function names, without having to go back and change basic parts of the code to implement new physics.

\section{Self-gravitating Magnetohydrodynamics}

In solving the equations of self-gravitating magnetohydrodynamics, we use a conservative formulation of hydrodynamics and a constrained transport approach to the magnetic induction equation. The physical meaning of a conservative equation is that (modulo source terms) the time rate of change of a conserved quantity in a volume is equal to a flux through the volume's enclosing surface. This meaning is built into our finite-difference representation of the hydrodynamics equations; divergences are represented in discrete correspondence to their mathematical definition, using zone volumes and face areas. Our discretization of the Poisson equation for the gravitational potential takes a similar form, since the Laplacian operator can be expressed as the divergence of a gradient. In our method of constrained transport, the curl of the electric field in the induction equation is represented in discrete correspondence to the mathematical definition of the curl, using zone face areas and edge lengths. The Stokes and divergence theorems then ensure a divergence-free magnetic field as required by Maxwell's equations.

Our solution methods are second order in space and time. Accurate computation of fluxes at zone faces and electric fields at zone edges is a key feature. We employ so-called `central schemes,' which are able to capture shocks with an accuracy comparable to Riemann solvers, but with much greater simplicity. We use a second-order Runge-Kutta time stepping algorithm, made possible by the semi-discrete formulation of the central scheme. In doing so we evolve all levels of the mesh synchronously, avoiding problems we encountered in self-gravitating systems with `asynchronous' evolution. Parallelization is achieved by giving each processor its share of spatial zones at each level of refinement. We rely on the PETSc library for the distributed matrix inversion required by the gravity solver.

\begin{figure}
\begin{center}
$\begin{array}{@{\hspace{-0.25in}}c@{\hspace{-0.75in}}c}
\includegraphics[width=4in]{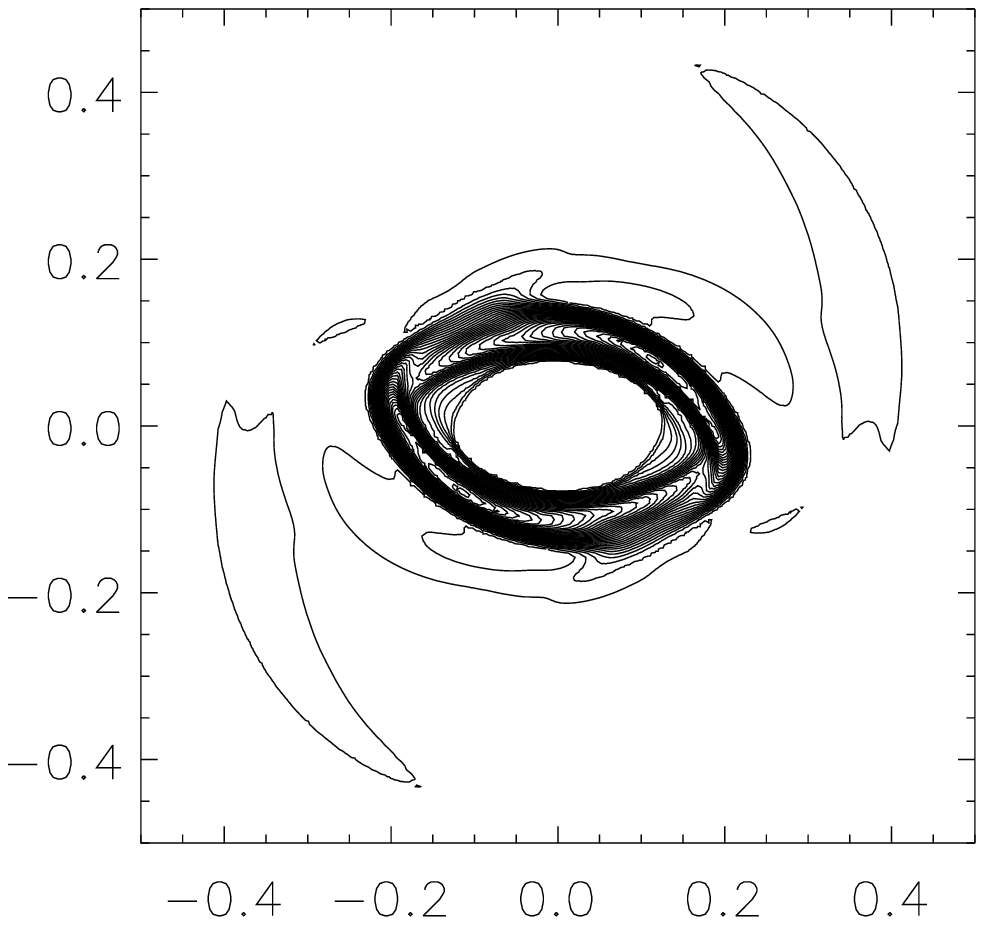} &
\includegraphics[width=4in]{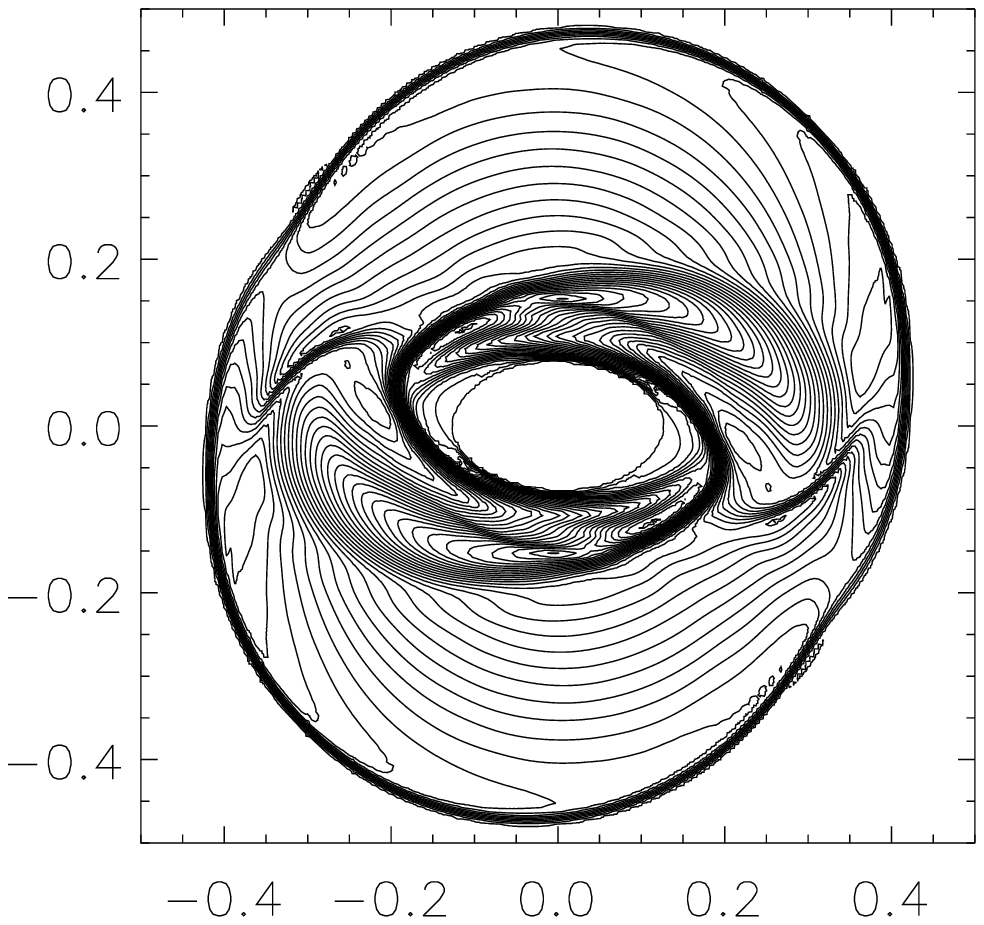} 
\end{array}$
\end{center}
\caption{Density (left panel) and thermal pressure (right panel) at $t=0.295$ for the second rotor problem given in Ref. \protect\cite{toth00}.  40 countours were used to produce the plots, with $0.512 \le mn \le 9.622$ and $0.010 \le p \le 0.776$.  A $200\times 200$ grid was used to produce the results.}
\label{rotorContours}
\end{figure}

We have been working with a number of hydrodynanic and magnetohydrodynamic test problems, one of which is shown here. 
The rotor problem---which consists of a rapidly rotating dense fluid, initially cylindrical, threaded by an initially uniform magnetic field---was devised to test the onset and propagation of strong torsional Alfv\'en waves into the ambient fluid. 
We have computed a version of the rotor problem with initial data identical to a so-called 'second rotor problem,' \cite{toth00} and display the results in figure \ref{rotorContours}.  

\section{Neutrino Transport}

Neutrino radiation transport is the greatest computational challenge in supernova simulations. Neutrino distributions must be tracked in order to compute the transfer of lepton number and energy between the neutrinos and the fluid. One challenge is constructing a formalism \cite{cardall03} and discretization that allows both energy and lepton number to be conserved to high precision. 

Two other challenges are associated with the limits of computational resources: the solution of a very large nonlinear system of equations \cite{cardall05b}, and neutrino interaction kernels of high dimensionality. A large system of nonlinear equations requiring inversion (as opposed to explicit updates) results from the disparity between hydrodynamic and particle interaction time scales, which motivates implicit time evolution. The nonlinear solve is achieved with the Newton-Raphson method. A fixed-point method employing a preconditioner that splits the space and momentum space couplings is used for the linear solve required within each Newton-Raphson iteration. Because neutrino interactions are expensive to compute on-the-fly, we have implemented interpolation tables. Particularly for neutrino scattering and pair interactions---which depend on neutrino states before and after the collision---the interaction kernels are of high dimensionality, requiring a globally distributed table. 

\section{Outlook}

We have made a promising start on {\em GenASiS,} a new code being developed to study the explosion mechanism of core-collapse supernovae. Our plan is to include all the relevant physics---including magnetohydrodynamics, gravity, and energy- and angle-dependent neutrino transport---in a code with adaptive mesh refinement in two and three spatial dimensions \cite{cardall05c}. Parallelization and implementation on the adaptive mesh are not yet complete, and the physics components have not yet been fully integrated; but steady progress and the successful completion of test problems give us confidence that we are well on our way towards a tool that will provide important insights into the supernova explosion mechanism.

\ack
We gratefully acknowledge S.~W. Bruenn's contribution of subroutines for the computation of neutrino interaction kernels. We  thank R.~D. Budiardja and M.~W. Guidry for discussions on the Poisson solver, and R.~D. Budiardja for helping with that solver's interface to the PETSc library. 
This work was supported 
by  Scientific Discovery Through
Advanced Computing (SciDAC), a program of the Office of Science of the U.S. Department of Energy (DoE); and by Oak Ridge National Laboratory, managed by UT-Battelle, LLC, for the DoE under contract DE-AC05-00OR22725.

\section*{References}

\def\jcp{{\em J. Comput. Phys.\ }}


\begin{thebibliography}{9}

\bibitem{mezzacappa05} Mezzacappa A 2005 {\em Annu. Rev. Nucl. Part. Sci.} in press

\bibitem{cardall05} Cardall C Y 2005 {\em Nucl. Phys. B (Proc. Suppl.)} in press ({\em Preprint} astro-ph/0502232)

\bibitem{toth00} T\'oth G 2000 \jcp {\bf161} 605

\bibitem{cardall03} Cardall C Y 2003 {\em Phys. Rev. D} {\bf 68} 023006 ({\em Preprint} astro-ph/0212460); Cardall C Y, Lentz E J, and Mezzacappa A 2005, {\em Phys. Rev. D} {\bf 72} 043007 ({\em Preprint} astro-ph/0510702)

\bibitem{cardall05b} Cardall C Y 2005 in {\em Numerical Methods for Multidimensional Radiative Transfer Problems} ed R Rannacher and R Wehrse ({\em Preprint} astro-ph/0404401)

\bibitem{cardall05c} Cardall C Y, Razoumov A, Endeve E, and Mezzacappa A 2005 in {\em Open Issues in Core Collapse Supernova Theory} ed A Mezzacappa and G M Fuller (Singapore: World Scientific) in press ({\em Preprint} astro-ph/0510704)

\end{thebibliography}
\end{document}